\documentclass[%
reprint,
 amsmath, amssymb, floatfix, aps, prapplied,longbibliography]{revtex4-2}

\usepackage{xcolor}
\usepackage{graphicx}% Include figure files
\usepackage{dcolumn}% Align table columns on decimal point
\usepackage{bm}% bold math
\usepackage{siunitx}% unit type setting
\usepackage{multirow}
\newcommand{\subs}[2]{{#1}_{\mathrm{#2}}}
\newcommand{\sups}[2]{{#1}^{\mathrm{#2}}}
\newcommand{\vect}[1]{\boldsymbol{#1}}

\newcommand{\mean}[1]{\langle #1 \rangle}
\newcommand{\kB}{\subs{k}{B}}
\newcommand{\params}{\vect{\theta}}
\newcommand{\Params}{\vect{\Theta}}
\newcommand{\indeps}{\vect{\phi}}
\newcommand{\pd}[2]{\partial #1 / \partial #2}
\newcommand{\PD}[2]{\frac{\partial #1}{\partial #2}}

\newcommand{\Frac}[2]{#1 / #2}

\begin{document}

\title{Weighing an optically trapped microsphere in thermal equilibrium with air}

\author{L.~E.~Hillberry}
 \email{lhillber@utexas.edu}
  \affiliation{%
    Center for Nonlinear Dynamics and Department of Physics,
    The University of Texas at Austin, Austin, TX 78712-1081, USA}%
\author{Y.~Xu}
 \affiliation{%
    Center for Nonlinear Dynamics and Department of Physics,
    The University of Texas at Austin, Austin, TX 78712-1081, USA}%
\author{S.~ Miki-Silva}
 \affiliation{%
    Center for Nonlinear Dynamics and Department of Physics,
    The University of Texas at Austin, Austin, TX 78712-1081, USA}%
\author{G.~H.~Alvarez}
 \affiliation{%
    Center for Nonlinear Dynamics and Department of Physics,
    The University of Texas at Austin, Austin, TX 78712-1081, USA}%
\author{J.~E.~Orenstein}
 \affiliation{%
    Center for Nonlinear Dynamics and Department of Physics,
    The University of Texas at Austin, Austin, TX 78712-1081, USA}%
\author{L.~C.~Ha.}
 \affiliation{%
    Center for Nonlinear Dynamics and Department of Physics,
    The University of Texas at Austin, Austin, TX 78712-1081, USA}%
\author{D.~S.~Ether}
 \affiliation{%
    Center for Nonlinear Dynamics and Department of Physics,
    The University of Texas at Austin, Austin, TX 78712-1081, USA}%
\author{M.~G.~Raizen}
% \email{raizen@physics.utexas.edu}
 \affiliation{%
    Center for Nonlinear Dynamics and Department of Physics,
    The University of Texas at Austin, Austin, TX 78712-1081, USA}%
\date{\today}
\begin{abstract}
We report a weighing metrology experiment of a single silica microsphere
optically trapped and immersed in air. Based on fluctuations about thermal
equilibrium, three different mass measurements are investigated, each
arising from one of two principle methods.  The first method is based on
spectral analysis and enables simultaneous extraction of various system
parameters. Additionally, the spectral method yields a mass measurement with
systematic relative uncertainty of 3.0\% in 3~s and statistical relative
uncertainty of 0.9\% across several trapping laser powers. Parameter values
learned from the spectral method serve as input, or a calibration step, for
the second method based on the equipartition theorem. The equipartition method
gives two additional mass measurements with systematic and statistical relative
uncertainties slightly larger than the ones obtained in the spectral method,
but over a time interval 10 times shorter. Our mass estimates, which are
obtained in a scenario of strong environmental coupling, have uncertainties
comparable to ones obtained in force-driven metrology experiments with
nanospheres in vacuum. Moreover, knowing the microsphere's mass accurately and
precisely will enable air-based sensing applications.
\end{abstract}
\maketitle
\section{\label{sec:intro} Introduction}
\par Optical trapping of nano- and micro-scale
objects~\cite{ashkin70,ashkin00,ashkin06} has become a
paradigmatic tool in diverse fields, from micro-manipulation of biological
samples~\cite{ashkin87a,ashkin87b,neuman08,pontes08,frases09,pontes11,moeendarbary14,nicholas14,ayala16,nussenzveig18,Nussenzveig19,bustamante09}
to center-of-mass cooling experiments~\cite{li10,gieseler12,gieseler20}
aiming to observe macroscopic quantum
effects~\cite{kaltenberg16,tebbenjohanns20,delic20}, to metrology
experiments~\cite{hebestreit18b,ricci19,zheng20} with optomechanical sensing
applications \cite{ranjit15,ranjit16,monteiro17,schnoering19,millen20}. In
such experiments, a tightly focused laser beam, named the optical
tweezer~\cite{ashkin86,ashkin11,gennerich17}, is used to polarize a dielectric
particle and harmonically confine it to the beam's intensity maximum.
\par It is often desirable to monitor the trapped particle's position
as a function of time, so a position-sensitive detector must be
calibrated. Calibrating the detector usually requires knowledge of the
trapped particle's mass~\cite{hebestreit18b}. However, $\mathrm{SiO_2}$
nano and microspheres, often the object-of-study in levitated optomechanics
experiments, do not have a readily-known mass. The St\"ober process used to
manufacture these particles \cite{stober68} yields very spherical results with
a low dispersion of radius ($\sim 3\%$), but a mass density which can vary in excess of
20\%~\cite{stober68,blakemore19}. Calculated with these values, the uncertainty
in mass is about 22\%. For this reason, recent work has focused on mass
metrology of nano and microspheres optically trapped in vacuum using methods
of electrostatic levitation~\cite{blakemore19}, oscillation~\cite{ricci19},
and trapping potential nonlinearities~\cite{zheng20}, and, most recently a
drop-recapture method perfomed in air~\cite{Carlse20}. The mass uncertainty
achieved in each of these experiments is at the level of one to a few percent.
Each has unique advantages like no assumptions on particle geometry,
and distinct challenges, e.g., control of the particle's charge, accurate
modelling of local potentials (gravitational, electric, or optical), or vacuum
capabilities including feedback cooling.
\begin{figure}[htp]
\includegraphics[width=3.375in]{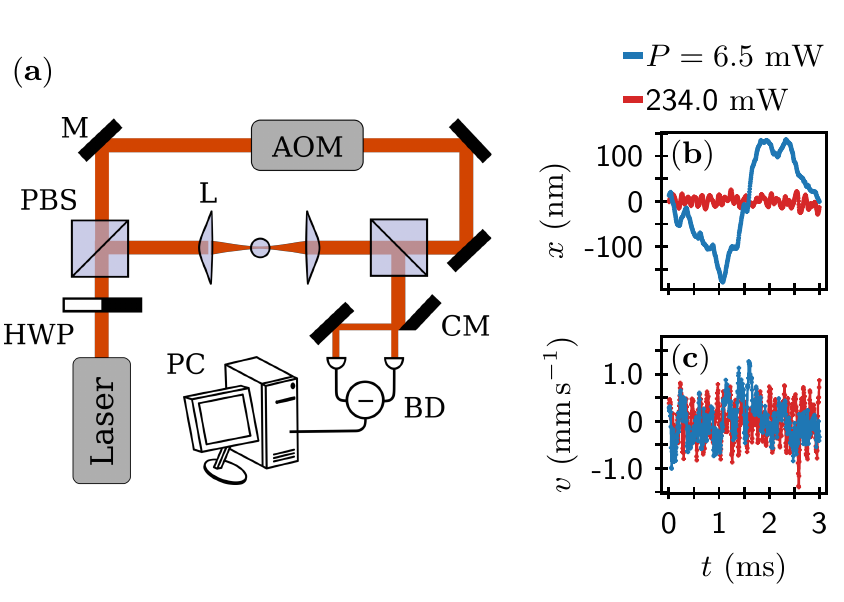}
\begin{centering}
    \caption{ \label{fig:fig1}
    (a) Schematic of the dual beam optical trap and the detection system:
        (M) dielectric mirror; (HWP) half wave plate; (PBS) polarizing
        beamsplitter; (AOM) acoustic optical modulator; (L) aspheric lense;
        (CM) cut mirror; (BD) balanced photodetector; (PC) personal computer.
    (b) An example position trace at high and low trapping power;
    (c) An example velocity trace which is computed with an $\rm
        8^{th}$-order-accuracy numeric finite difference of the
        data in (b).
    }
\end{centering}
\end{figure}
\par Here, we report on a mass metrology experiment with uncertainty
similar to previous work, but performed on a $1.5~\si{\micro m}$ radius
$\rm SiO_2$ microsphere optically trapped in air \cite{gong18} at room
temperature and pressure.  Our experiment employs a dual-beam optical
trap~\cite{ashkin70,vanderhorst08}, sketched in Fig.~\ref{fig:fig1}(a)
and elaborated upon in
Our system remains in thermal equilibrium at all times making for a simple
protocol. Moreover, we explore two distinct methodologies leveraging our
detector's high spatiotemporal resolution.  In the first \emph{spectral
method}, we fit an average voltage signal power spectral density (PSD) to
simultaneously extract parameters which make no assumptions on the physical
conditions of the experiment. We then fix conditions known with high accuracy
--- the air temperature, air viscosity, and particle radius --- to compute
the harmonic trap strength $k$, microsphere mass density $\rho$, and detector
calibration factor $\beta$, as well as the uncertainties and correlations of
these parameters.  The microsphere mass is similarly calculated by combining
fitting and fixed parameters.
\par In the second \emph{equipartition method}, we compute the voltage signal
and voltage-derivative signal variances from which we deduce the particle mass
in two additional ways. Doing so requires a detector with sufficient resolution
to observe the particle's instantaneous velocity, pioneered in ~\cite{li10,
kheifets14}.  The equipartition methods additionally require knowledge of
either the harmonic trap strength $k$ or calibration factor $\beta$ which must
be determined via the spectral method first. The spectral method demands a high
volume of data to sufficiently smooth the experimental PSD, and is in that
sense slow. The equipartition methods, once an initial spectral calibration
is performed, require 10 times less data to achieve similar uncertainty in
subsequent mass measurements.
\par Making precise and fast mass measurements in a system strongly coupled
to the environment might have applications in scenarios where either the mass
changes with time but temperature is fixed, for instance in heterogeneous
nucleation \cite{ishizaka11,krieger12}, or the mass is fixed but temperature
changes, like in Rayleigh-B\'enard convection~\cite{chilla12,iyer20}.
\par This Article is organized as follows: first, we review the
relevant physics and outline our PSD parameter estimation method in
Section~\ref{sec:params}. In Section~\ref{sec:mass} we show how the PSD
parameters, along with the equipartition theorem, allow us to weigh the
microsphere in three different ways. There, we also present the results which
we further discussion in Section \ref{sec:discussion}. Finally, we conclude
with this work's significance in section \ref{sec:conclusion}.
\section{\label{sec:params} Power spectral density parameter estimation}
\par The dynamics of the trapped microsphere along the $x$-axis may be modeled
by the harmonically bound Langevin equation of motion
\begin{equation}
    m \ddot{x} + \gamma \dot{x} + k x = F(t),
\label{eq:eom}
\end{equation}
where $m = 4 \pi \rho R^3/3$ is the mass of the microsphere with radius $R$
and density $\rho$, $\gamma = 6 \pi \eta R$ is the Stokes friction coefficient,
$\eta$ is the viscosity of air, and $k$ is the trap strength. The stochastic
thermal force $F(t) = g \xi(t)$ is assumed to have the form of zero-mean
$\mean{\xi(t)} = 0$, delta-correlated $\mean{\xi(t) \xi(t')} = \delta(t-t')$
white noise with strength $g=\sqrt{2 \kB T \gamma}$ (according to the
fluctuation-dissipation theorem), and in which $T$ is the air temperature,
$\kB$ is Boltzmann's constant, and $\mean{\cdot}$ denotes ensemble averages
over realizations of $\xi$. Writing Eq.~\eqref{eq:eom}, in terms of the Fourier
transforms $\tilde{x}(\omega)$ and $\tilde{F}(\omega)$
\footnote{The Fourier integrals are defined as
    $x(t) = \frac{1}{2\pi} \int_{-\infty}^{\infty} \tilde{x}(\omega)
        e^{-\imath\omega t} \mathrm{d} \omega$ and
    $\tilde{x}(\omega) = \int_{-\infty}^{\infty}x(t^{\prime})
        e^{\imath\omega t^{\prime}} \mathrm{d} t^{\prime}$.
}
lets one deduce the position PSD $S_x(\omega)$ such that
$\mean{\tilde{x}(\omega)\tilde{x}(\omega^{\prime})} =
    S_x(\omega)\delta(\omega-\omega^{\prime})$
\cite{mandel95}, where $\omega=2 \pi f$ is the angular frequency.
\par In our experiment, we record a unitless voltage signal $V(t) = V_{-}(t)
/ V_{+}(0)$, where $V_{-}(t)$ is proportional to the difference in optical
power delivered to the two ports of the balanced photodetector at time
$t$ and $V_{+}(0)$ is proportional to the total detection power at time
$t=0$. Normalizing the signal in this way accounts for small variations in
detected power upon changing the trapping laser power. $V(t)$ is proportional
to the microsphere's position along the $x$-axis: $x(t) = V(t) / \beta$, where
$\beta$  is the calibration factor which we report in units of $\si{\micro
m^{-1}}$.
\par From such considerations, the theoretical (one-sided) PSD of our voltage
signal is understood to be
\begin{equation}
    S_V(\omega) =
        \beta^2 \frac{4 \kB T \gamma}{(m \omega^2 - k)^2 + \gamma^2 \omega^2} \,.
\label{eq:psdfull}
\end{equation}
Multiple trials of experimental power spectra must be averaged together
before we attempt to learn relevant physical parameters. We collect
10 trials of the voltage signal, each 0.3 seconds long, at a sampling
rate of 50 MHz. In post-processing, the signal is low-pass filtered by
averaging together non-overlapping blocks of 256 samples for improved spatial
resolution. The new effective sampling rate is 195 kHz. Using \emph{Bartlett's
method}~\cite{oppenheim01} with four windows per trial --- for a
total of 40 averages of length $\mathcal{T} = \si{84~ms}$ --- we estimate the
experimental voltage PSD, denoted $\hat{S}_{V,k} = \hat{S}_V(f_k)$. The index
$k$ labels the descrete frequencies at which the experimental PSD is known. The
frequency resolution is $f_{k+1} - f_k = \mathcal{T}^{-1}$.
\par Once a set of trials is collected, we fit the experimental data
$\hat{S}_V$ to
\begin{equation}
    S_V(f; \params) = \frac{1}{a + b f^2 +c f^4},
\label{eq:psdfit}
\end{equation}
in which we have defined the column vector of free parameters $\params=
\sups{(a,b,c)}{T}$. In particular, the fit is done using the maximum likelihood
estimation method~\cite{cole14,norrelykke10,dawson19} which we briefly outline
next.
\par First, note that each data point of an $n$-trial-averaged PSD is
subject to gamma-distributed noise (the convolution of $n$ exponential
distributions)~\cite{berg-sorensen04,norrelykke10}, written
\begin{equation}
    \mathcal{P}(\hat{{S}}_{V,k}) = \frac{1}{S_{V,k}} \frac{n^n}{\Gamma(n)}
        \left ( \frac{\hat{{S}}_{V,k}}{S_{V,k}} \right )^{n-1}
         \exp \left (-n \frac{ \hat{{S}}_{V,k}}{S_{V,k}} \right )\, ,
     \label{eq:gamma}
\end{equation}
where $\Gamma(n) =(n-1)!$ is the gamma function and $S_{V,k}$ is the mean
value of the distribution.  Then, the likelihood of measuring the entire data
set data $\hat{S}_V$ given a model $S_{V,k} = S_V(f_k,\params)$ is the joint
distribution
\begin{equation}
    \mathcal{P}(\hat{S}_V | \params) =
        \prod_k \mathcal{P}(\hat{S}_{V,k}).
    \label{eq:likelihood}
\end{equation}
Maximizing the likelihood~\eqref{eq:likelihood} is equivalent to minimizing
the negative-log-likelihood
\begin{equation}
    \mathcal{L}(\params, \hat{S}_V) =
        n \sum_k \left(\log[S_V(f_k; \params)] +
            \frac{\hat{S}_V(f_k)}{S_V(f_k; \params)}\right) + C \, ,
\label{eq:objective}
\end{equation}
where $C = \sum_k[\log\Gamma(n)- n \log n-(n-1) \log \hat{S}_V(k)]$ is a
constant with respect to the free parameters and thus inconsequential for the
minimization, and $n=40$ is the number of spectra averaged together in the
experiment.
\par
\begin{figure*}[!htbp]
\includegraphics[width=5.0625in]{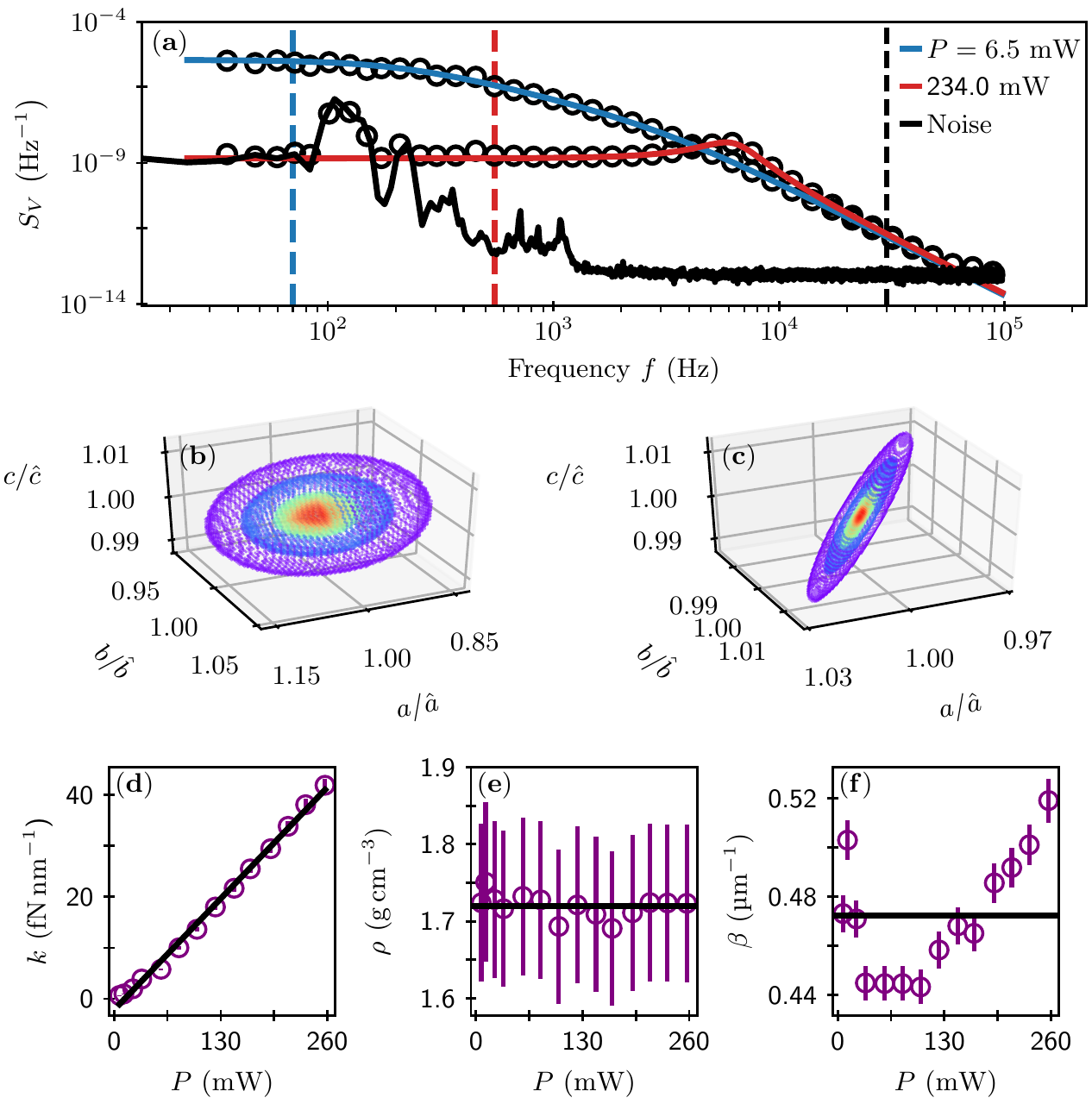}
\caption{\label{fig:fig2}
    (a) Voltage PSD for 6.5 mW (blue) and 234.0 mW (red) of trapping laser
        power. Experimental data is depicted with open circles and consists of
        40 independent PSD averages and further bin-averaged on a logarithmic
        horizontal scale for visualization purposes. The solid lines are the
        maximum-likelihood best fit to eq.~\eqref{eq:psdfit}. The vertical
        dashed lines mark the bounds on the data used in the fit. The lower
        bound is color coded with the fitting line and the upper bound (black
        dashed line) is shared. The noise spectrum (solid black) was collected
        under identical detection conditions as the other two curves but
        with no microsphere present. We see a technical noise floor at high
        frequencies and electronic/laser noise, including 60 Hz harmonics and
        $1/f$ noise, at low frequencies. The noise peak at around 120 Hz must
        be omitted when fitting the 234.0 mW spectrum, as indicated by the
        lower bound (red dashed line).
    (b) For 6.5 mW of laser power, we plot the isosurfaces of $\mathcal{P} =
        \exp(-\mathcal{L})$ (see eq.~\eqref{eq:objective}) as a function of
        fitting parameters $\params$. The isosurfaces are taken at Gaussian
        widths of 3-sigma (purple), 2-sigma (blue), and a core from the peak
        (red) to the 1-sigma-width (green).
    (c) As in (b) but for 234.0 mW of trapping laser power. By fitting the the
        likelihood data clouds shown in (b) and (c) to a three-variate Gaussian
        distribution, we extract the best fit parameters $\hat{\params}$ and
        the variance-covariance matrix $\vect{\Sigma}_{\params}$.
    (d) Trap strength $k$,
    (e) microsphere mass density $\rho$, and
    (f) calibration constant $\beta$, each extracted from PSD fits as a
        function of laser power $P$. In (d-f), error bars reflect systematic
        uncertainty calculated by error propagation including correlation among
        fitted parameters.
        }
\end{figure*}

 Good starting values for the minimization can be
calculated analytically and implemented numerically, a convenient
feature which is not possible if one attempts to fit directly to
eq.~\eqref{eq:psdfull}~\cite{norrelykke10}. Maximum likelihood fitting
accounts for the gamma distributed PSD data, unlike more common least-squares
fitting algorithms which assume normally-distributed noise and thus provide
biased PSD parameter estimations~\cite{norrelykke10}. In the end, the
minimization gives the best fit parameters $\hat{\params} = \sups{(\hat{a},
\hat{b}, \hat{c})}{T}$ which maximize the likelihood of the data given the
model $\mathcal{P}\left(\hat{S}_V \big \vert \params \right) = \exp \left
[-\mathcal{L}(\params, \hat{S}_V) \right ]$. In Fig.~\ref{fig:fig2}(a) we show
experimental PSD and the best-fit curve for two different trapping laser powers
and compare to the noise inherent to the detection system.
\par To measure the parameter fitting uncertainty and correlation, and
inspired by the \emph{profile likelihood} method \cite{dawson19,cole14}, we
scan $\params$ in the vicinity of $\hat{\params}$ over a volume of parameter
space to build up a three-variate probability distribution $\mathcal{P}$
(See Fig.~\ref{fig:fig2}(b)-(c)) which is fit to a three-variate Gaussian
distribution
\begin{equation}
\subs{\mathcal{P}}{G}(\params; \hat{\params}, \vect{\Sigma}_{\params})= \exp \left [ -\frac{1}{2}\sups{(\params - \hat{\params})}{T}\vect{\Sigma}_{\params}^{-1}(\params - \hat{\params})\right].
\end{equation}
\par The absolute residuals $\lvert \mathcal{P} - \subs{\mathcal{P}}{G} \rvert$
are bound below the 1\% level. The $\sups{95}{th}$ percentile is bound below
the 0.1\% level (see Appendix~\ref{sec:uncertainty}). The vector
$\hat{\params}$ resulting from the fit is taken as the best-fit parameter set.
The matrix $\vect{\Sigma}_{\params}$ resulting from the fit provides the
variance-covariance matrix of the fitted parameters:
\[
    \vect{\Sigma}_{\vect{\theta}} =
              \begin{pmatrix}
                \sigma^2_a & \sigma^2_{ab} & \sigma^2_{ac} \\
                \sigma^2_{ab} & \sigma^2_b & \sigma^2_{bc} \\
                \sigma^2_{ac} & \sigma^2_{bc} & \sigma^2_c \\
               \end{pmatrix}.
\]
The uncertainty in parameter $i$ is $\sigma_i = \left (
[\Sigma_{\params}]_{i,i} \right )^{1/2}$ and the correlation coefficient
between parameters $i$ and $j$ is $r_{i,j} = [\Sigma_{\params}]_{i,j} /
(\sigma_i \sigma_j)$, for $i,j=a,b,c$ (see Appendix~\ref{sec:uncertainty}) for
a visualization)
\par The fitting parameters $\params$ may be used to deduce a more physical set
of parameters: trap strength $k$, microsphere density $\rho$, and calibration
constant $\beta$. Each of the physical parameters $\Params = \sups{(k, \rho,
\beta)}{T}$ are a function of the fitting parameters and constant parameters
$R$, $T$, and $\eta$. That is, $\Params = \Params(\indeps)$, where we have
defined the vector of independent variables $\indeps=\sups{(\params, R, \eta,
T)}{T}$ (explicit formulae in Appendix~\ref{sec:parameters}). We now turn to the uncertainty
analysis of the constant parameters.
\par The microsphere radius is known to be $R=1.51~\si{\micro m}$ up to 3.0\%
uncertainty based on statistical analysis of $\sim 200$ microspheres imaged
with a scanning electron microscope~\cite{particlesizer}. Similar image analysis suggests $\epsilon
= a/b - 1$ to be 0.027, where $a/b \geq 1$ is the aspect ratio of the imaged
microspheres.  To first order in $\epsilon$, we estimate corrections to
the Stokes friction coefficient due to aspherical geometry~\cite{brenner64}
to be less than 1\%. Similar estimations apply to the microsphere volume,
so uncertainty in the radius dominates uncertainty in the geometry. The
air temperature, measured with a thermocouple before each trial, was found
to vary less than 0.05\% over the entire experimental run. The viscosity
of air, calculated as a function of temperature with Sutherland's model,
is found to vary over a similarly small range~\cite{chapman90}. Sutherland's
model is known to interpolate experimental viscosity data near room
temperature with an uncertainty below 0.09\% including effects of up to
10\% humidity~\cite{mulholland06}. Since the experiment is performed at
atmospheric pressure, the particle-environment interaction is outside the
Knudsen regime~\cite{millen14}. As a result, no laser-induced heating of the
microsphere is expected, and so thermal equilibrium is assumed.
\par In light of these observations, the variance-covariance matrix of fitting
and constant parameters may be approximated in the block-diagonal form
$\mathrm{diag}(\Sigma_{\theta}, \sigma^2_R, 0, 0)$ where the last two zeros
reflect the small relative uncertainty in $T$ and $\eta$ compared to that in
$a$, $b$, $c$, and $R$. The block diagonal form assumes correlation exists only
between the fitted parameters.
\par We calculate the variance-covariance matrix of the physical
parameters in terms of the fitting and constant parameters via the error
propagation equation~\cite{tellinghuisen01} $\vect{\Sigma}_{\Params}
= \vect{J}_{\Params}\vect{\Sigma}_{\indeps}\vect{J}_{\Params}^{\rm
T}$. The Jacobian matrix (evaluated at the optimal fitting parameters)
is $(J_{\Params})_{i,j} =[\partial \Theta_i / \partial \phi_j]_{\params =
\hat{\params}}$. We have verified that the parameters and uncertainties
deduced by the procedure described here and conveniently visualized in
Fig.~\ref{fig:fig2}(b)-(c) agree quantitatively with the Monte-Carlo method
which generates and fits many artificial PSDs by sampling the appropriate
gamma distribution. Our technique yields directly the probability density,
sidestepping the need for binning and fitting or kernal-density estimating the
Monte-Carlo results.
\par We now understand how to estimate $k$, $\rho$, and $\beta$, including
uncertainty and correlation, from an experimental voltage PSD $\hat{S}_V$. The
results are presented in Fig.~\ref{fig:fig2}(d)-(f) for experiments on
the same trapped microsphere and which scan the trapping laser power
from 6.5 to 257.2 mW. We observe no unexpected dependence of the physical
parameters on laser power except for the calibration constant which exhibits
a non-monotonic curve, first decreasing then increasing with laser power
(Fig.~\ref{fig:fig2}(d)). Thus, we conclude heating of the microsphere due
to the laser is inconsequential because of the strong environmental coupling.
This is not the case for experiments in vacuum~\cite{millen14}.  The trend
in $\beta$ is reproducible when the experiment is repeated with different
microspheres, suggesting the source is most likely slight beam deviations
caused by the half wave plate/polarizing beamsplitter pairs used to control the
trapping and detected power.
\section{\label{sec:mass} Mass measurement technique}
\par Upon learning PSD fitting parameters, presented in
section~\ref{sec:params}, it is straightforward to estimate the
mass using the density and radius of the microsphere. However, the
equipartition theorem, $\kB T = m \langle \dot{x}^2 \rangle = k \langle
x^2 \rangle$, provides two additional possibilities.  The three mass
measurements written in terms of the augmented independent variables
$\indeps^{\prime}=\sups{(a,b,c,\mean{\dot{V}^2}, \mean{V^2}, R, \eta, T)}{T}$
read
\begin{gather}
    m_1(\indeps^{\prime}) = \frac{4}{3} \pi R^3 \rho  \, ,\\
    m_2(\indeps^{\prime}) = \frac{\kB T}{\mean{\dot{V}^2}}\beta^2 \, ,\\
    m_3(\indeps^{\prime}) = \frac{\mean{V^2}}{\mean{\dot{V}^2}} k \, .
\end{gather}
\par
\begin{figure}[htp]
\includegraphics[width=3.375in]{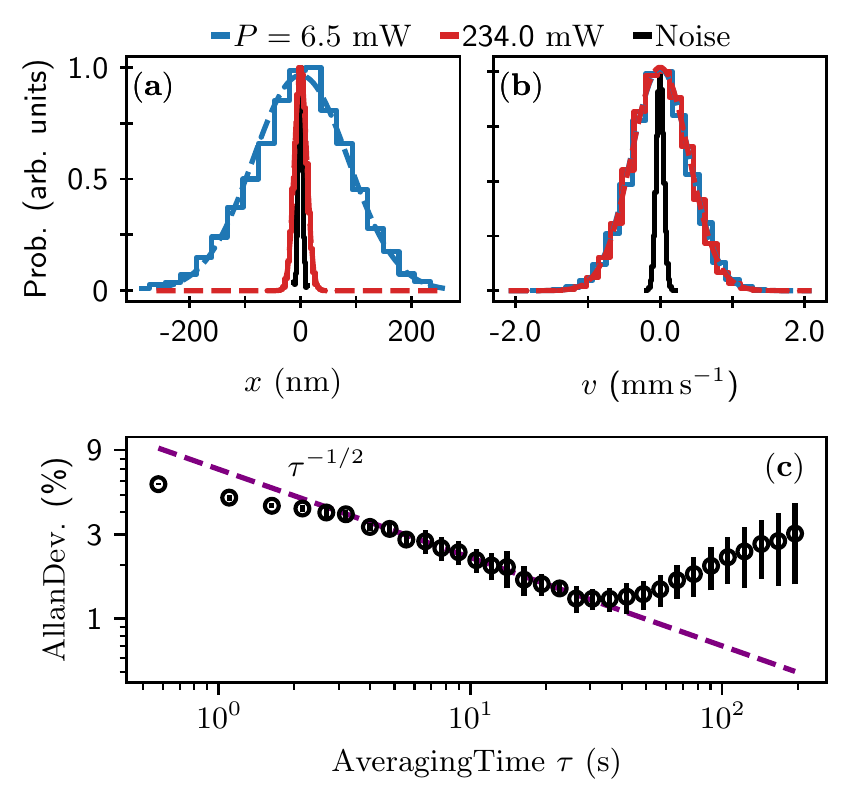}
\caption{\label{fig:fig3}
    (a) Position probability density (solid steps) of the microsphere's
        position over a single 0.3 second trial for 6.5 mW (blue) and 234.0 mW
        (red) of trapping laser power. Dashed lines correspond to a Gaussian
        fit with variance as the only free parameter. The black line is
        a histogram of the signal when no microsphere is present.
    (b) Velocity probability density. Colors are shared with (a). As expected
        by the equipartition theorem, the width of the position probability
        density gets narrower with increasing laser power while the width
        velocity probability density remains constant. All curves in (a) and
        (b) are normalized by their maximum value.
    (c) The relative Allan deviation of variance calculated from a 14 minute
        voltage signal decreases with increasing averaging time according to
        a $-1/2$ power law. The decay trend continues until a minimum of about
        1\% is reached in 30 seconds. Normalization is provided by the value
        of variance corresponding to the minimum Allan deviation.  Error bars
        reflect the standard deviation of three trials.}
\end{figure}

 The benefit of $m_2$ and $m_3$ is that, once a PSD fit is used to
calibrate the system, further data can be collected to estimate the variances
$\mean{\dot{V}^2}$ and $\mean{V^2}$, which may be used to update the mass
measurement in the case it changes with time. Of course, there is nothing to
update if the mass is unchanging. Nonetheless, to make use of methods $m_2$
or $m_3$, we must make an adequate estimate of the required variances. In
Fig.~\ref{fig:fig3}(a)-(b) we show the histograms of position and velocity
(proportional to $V$ and $\dot{V}$, respectively) for high and low trapping
laser power. The histograms consists of data from a single 0.3 second
trial. Overlaid on each histogram are Gaussian fits with variance as the
only free parameter. Uncertainty in the variance calculation is taken as the
standard deviation of variances calculated across ten trials.
\par For an uncorrelated voltage trace of length $\tau$, the
uncertainty in the variance estimate scales as $\tau^{-1/2}$, which is a
thermally-limited trend. However, at short times ($\tau<m / \gamma$), the
data is correlated due to the microsphere's dynamics and, at long times,
slow drifts in the system tend to affect the signal's variance. One way to
quantify those correlations and to determine the optimal time over which
our measurements are thermally limited is performing an Allan-deviation
stability analysis~\cite{allan66,czerwinski09,hebestreit18b,schnoering19}.
Figure~\ref{fig:fig3}(c) shows the results of our Allan-deviation experiment
performed with 22.8 mW of trapping laser power. Accordingly, our system is
stable out to about 30~s, so using 0.3~s of data for estimating the variances
allows for 100 independent mass measurements before the slow drifts demand
recalibration of the apparatus. It is in this sense that methods $m_2$ and
$m_3$ are faster than $m_1$.
\begin{figure}[!htp]
\includegraphics[width=3.375in]{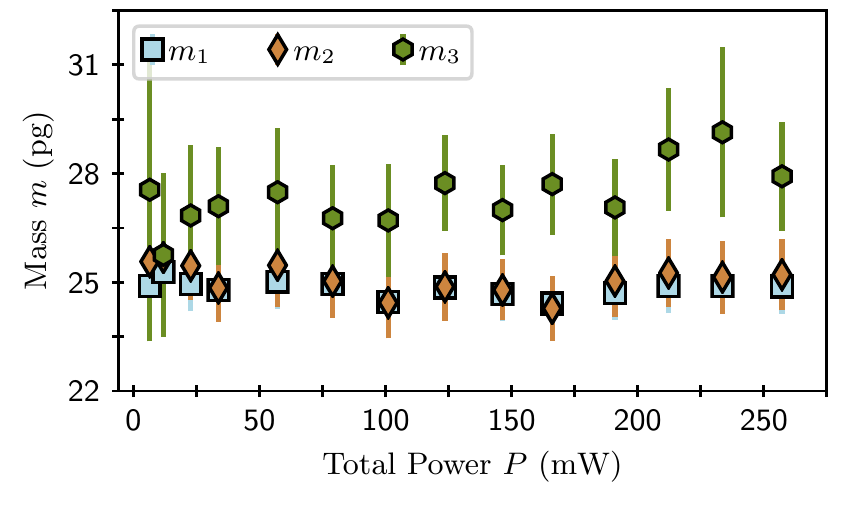}
\caption{\label{fig:fig4}
    Mass measurement comparison: blue squares, orange diamonds, and
    green hexagons denote, respectively, $m_1$, $m_2$, and $m_{3}$.
    mass measurements as the total trapping power $P$ is varied. Error
    bars denote the systematic uncertainties $\sups{\sigma}{sys.}_{m_i}$,
    $i=1,2,3$. Statistical uncertainties $\sups{\sigma}{stat.}_{m_i}$ are
    calculated as standard deviations across the ensemble of mass estimates at
    different laser powers.}
\end{figure}
\par In Fig.~\ref{fig:fig4} we show the results of our three mass measurement
procedures. We find $\bar{m}_1 = 24.8~\si{pg}$, $\bar{m}_2 = 25.1~\si{pg}$,
and $\bar{m}_3 = 27.4~\si{pg}$ where the over bar denotes an average over the
14 experiments at different total trapping laser powers. Error bars reported
are calculated by considering covariance of the PSD fitting parameters,
and uncertainties in both fixed parameters and variances. We consider such
error bars systematic uncertainty, denoted $\sups{\sigma}{sys.}_{m_i}$,
$i=1,2,3$. The statistical uncertainty (or fluctuation), denoted
$\sups{\sigma}{stat.}_{m_i}$, is calculated as standard deviation across
the 14 experiments at different laser powers. Measurement $m_1$, which is
based entirely on the PSD analysis, has the smallest relative error bars
($\sups{\bar{\sigma}}{sys.}_{m_1} / \bar{m}_1= 3.0 \%$) and the smallest
relative statistical uncertainty ($\sups{\sigma}{stat.}_{m_1} / \bar{m}_1 =
0.9\%$). Measurement $m_2$, which supplements the PSD analysis by estimating
the voltage-derivative variance, agrees well with $m_1$, albeit with relative
error bars at 4.1\% and relative statistical fluctuations at 1.6\%. The
benefit of $m_2$ is that, once an initial PSD analysis is performed,
parameters like mass (or temperature) can be subsequently updated 10 times
faster than collecting data for additional PSD analysis. Measurement $m_3$ has
the largest systematic and statistical relative uncertainties, 6.7\% and 3.0\%
respectively. Furthermore, method $m_3$ displays additional systematic error
because it deviates from $m_1$ and $m_2$ by nearly 10\%. We speculate on the
source of this uncertainty in the next section.
\par For comparison, the mass according to the manufacturer
values of density $\subs{\rho}{Bangs}=2.0~\si{g~cm^{-3}}$
($\sigma_{\subs{\rho}{Bangs}}/\subs{\rho}{Bangs}=20\%$) and our radius
measurement $R=1.51~\si{\micro m}$, $(\sigma_R/R=2.9\%)$ is $\subs{m}{Bangs} =
28.84~\si{pg}$ with an uncertainty of 22\%  which agrees within the uncertainty
tolerance of all our mass measurements, despite the discrepancy of the mean
values.
\section{\label{sec:discussion} Discussion}
\par The systematic bias in $m_3$ of Fig.~\ref{fig:fig4} is hypothesized
to be dominated by the low frequency electronic noise apparent in
Fig.~\ref{fig:fig2}(a). By selecting an appropriate lower bound for the fit,
the spectral method easily removes the influence of the noise resonances,
the most severe of which appears at 120 Hz with a width of about 100
Hz. However, the time domain estimate of $\mean{V^2}$ includes variance
due to that noise. To estimate the effects of such noise, we can model the
experimental PSD as the sum of the best fit PSD and the experimental noise
PSD containing the noise peak near 120 Hz, $S_V(f)=\sups{S}{best}_V(f) +
\sups{S}{noise}_V(f)$. Using Parseval's theorem, $\langle q^2 \rangle =
\int_0^{\infty} S_q(f) \mathrm{d}f$, we have
\begin{equation}
    \mean{V^2} \approx
        \subs{\mean{V^2}}{PSD} +
            \int_{70~\si{Hz}}^{170~\si{Hz}} \sups{S}{noise}_V(f) \mathrm{d}f \, ,
\end{equation}
where  $\mean{V^2}$ is the variance of the voltage signal observed in the time
domain, $\subs{\mean{V^2}}{PSD} = \kB T \beta^2 / k$ is the variance estimate
provided by the PSD parameters. The excess variance $\Delta \mean{V^2}
\equiv \lvert \mean{V^2} - \subs{\mean{V^2}}{PSD} \rvert$ is about 10\%
of $\subs{\mean{V^2}}{PSD}$ , which agrees with the discrepancy between
$m_3$ and and the other measurements. The quantity $\Delta \mean{V^2}$ can
also be estimated by numerically integrating the observed noise spectrum.
We find that the integral of the noise PSD in frequency band from
70 Hz to 170 Hz predicts the observed excess variance and hence also the bias
in $m_3$.
\par The effects of low frequency noise resonances are suppressed when
estimating $\mean{\dot{V}^2}$. The reason is because, in general, $S_{\dot{q}}
\left(f\right)= (2 \pi f)^2 S_q(f)$, so high frequency components of a signal
have a quadratically larger weighting factor in the variance compared to
low frequency noise. We find, by direct calculation on our data, $\Delta
\mean{\dot{V}^2} \sim 2\%$, which agrees with numeric integration of
$\sups{S}{noise}_{\dot{V}}(f)$ over all frequencies above 80 kHz.
\par A recent experimental effort~\cite{ricci19} measured the mass of
$0.143~\si{\micro m}$ radius $\mathrm{SiO_2}$ spheres optically trapped in
vacuum to be 4.01 fg with 2.8\% uncertainty with 40~s of position data. Their
oscillating electric field method makes no assumption on particle shape
or density, though a density of $2.2~\si{g~cm^{-3}}$ agrees with their
measurements. In~\cite{blakemore19}, a $2.6~\si{\micro m}$ radius sphere is
optically trapped and levitated with a static electric field as the trapping
laser power is reduced, resulting in a mass measurement of 84 pg with
1.8\% uncertainty with 42 minutes of data.  The density is also measured
to be 1.55~\si{g~cm^{-3}} with 5.16\% uncertainty.  A third strategy
used in~\cite{zheng20} stabilizes oscillations of a $0.082~\si{\micro
m}$ radius sphere in the nonlinear-trapping regime to deduce the detector
calibration constant with 1.0\% uncertainty and a mass of 3.63 fg with 2.2\%
uncertainty. Finally, very recent work~\cite{Carlse20} used a drop-recapture
method and camera-based detection with time resolution that could not
quite resolve the microsphere's instantaneous velocity. Fitting position
autocorrelation functions, they measure their resin particle's radius to be
2.3 \si{\micro m} with 4.3\% statistical uncertainty. In the drop-recapture
experiments, 90 s worth of trials are used to deduce a mass of 55.8 pg
with 1.4\% statistical uncertainty and 13\% systematic uncertainty. The
authors combine the radius and mass measurements to deduce a density of
1.1 \si{g~cm^{-3}} with 9.1\% statistical uncertainty.
\par As a comparison, we present a summary of our physical parameter values
and uncertainties in Table~\ref{tab1}. Based entirely on thermal equilibrium
analysis, our two most accurate mass estimates have uncertainties of 3\% to 4\%
as compared to the 1\% to 2\% uncertainty in vacuum-dependent and 13\% in the
air-based, nonequilibrium methods. Further, all of our measurements are made
with significantly less position data. Interestingly, our density measurement
has comparable accuracy to the recent body of work using $\mathrm{SiO_2}$
particles, all of which sourced particles from the same manufacturer. The
variability and apparent radius dependence of measured density values
underscores the parameter's uncertainty inherint to the manufacturing process.
\par Most of the existing mass measurement methods are demonstrated
in a high vacuum environment where the experimental goals often
center around ground-state cooling or exceptionally sensitive force
transduction. Additionally, the existing methods rely on forces external to the
trap, often driving the system out of equilibrium and limiting their utility
as environmental sensors. Our method has the advantages of speed in that
between $10\times$ and $100\times$ less data is required compared to other
methods; environmental coupling, which unlocks future sensing applications;
and simplicity in that no additional experimental set up is required beyond
trapping and monitoring.
\par Disadvantages include the requirements of environmental coupling,
enough spatiotemporal resolution to resolve the instantaneous velocity, and
accurate knowledge of the particle radius. While an advantage for future
applications, environmental coupling critically limits heating effects of
the trapping laser and enables fast equilibration with the environment,
so our method will face complications in vacuum-based experiments. Accurate
heating/damping models and longer data traces could possibly overcome such
concerns. Instantaneous velocity resolution enables our fastest measurement,
$m_2$, but can be much more difficult in a liquid environment, though certainly
possible~\cite{kheifets14}. Accurate knowledge of the trapped particle
geometry was quantified statistically in our experiment, but less uniform
samples could significantly alter the error analysis.  In these cases, in situ
measurements of the trapped particle with optical microscopy, light scattering,
or autocorrelation function analysis could improve the error
budget.
\begin{table}[htp]
\caption{\label{tab1}
    Table of values and uncertainties. Reported values are the average over the
    power scan experiment, except for $k$ and $\mean{V^2}$ for which we report
    the range since these quantities scale linearly with $P$.  Also reported
    are the relative systematic uncertainties (sys.) averaged over the power
    scan experiment and the statistical uncertainties (stat.) which express the
    relative standard deviation over the power scan, where applicable.}
\begin{tabular}{ | c | c |d  d | c |}
 \hline
      \multirow{2}{*}{\textbf{Quantity}} & \multirow{2}{*}{\textbf{Value}} & \multicolumn{2}{c|}{\textbf{Uncertainty} (\%)} & \multirow{2}{*}{\textbf{Unit}} \\
      & & \multicolumn{1}{c}{~\text{sys.}} & \multicolumn{1}{c|}{\text{stat.}} & \\
 \hline
    $R$ & 1.51 & 2.9 & \multicolumn{1}{c|}{-} & \si{\micro m}\\
 \hline
    $\eta$ & 18.295 & 0.04 & \multicolumn{1}{c|}{-} & \si{\micro Pa~s}\\
 \hline
    $T$ & 295.50 & 0.05 & \multicolumn{1}{c|}{-} & \si{K}\\
 \hline
    $\mean{\dot{V}^2}\times 10^3$ & 36.4 & 2.4 & \multicolumn{1}{c|}{-} & \si{\micro s^{-1}}\\
 \hline
    $\mean{V^2}\times 10^3$ & (0.03, 1.53) & 8.9 & \multicolumn{1}{c|}{-} & arb. units\\
 \hline
    $k$ & (0.66, 49.1) & 3.1 & \multicolumn{1}{c|}{-} & \si{fN~nm^{-1}}\\
 \hline
    $\rho$ &  1.72 & 5.9 & 0.9 & \si{g~cm^{-3}} \\
 \hline
    $\beta$ & 0.47 & 1.6 & 5.0 & \si{\micro m^{-1}}\\
 \hline
    $m_1$ & 24.8 & 3.0 & 0.9 & \si{pg}\\
 \hline
    $m_2$ & 25.1 & 4.1 & 1.6 & \si{pg}\\
 \hline
    $m_3$ & 27.4 & 6.7 & 3.0 & \si{pg}\\
 \hline
\end{tabular}
\end{table}
\section{\label{sec:conclusion} Conclusions}
\par We have explored spectral and equipartition methods by which to measure
an optically trapped microsphere's mass while in thermal equilibrium with air.
With the former, we accurately extract physical parameters of trap strength
$k$, microsphere density $\rho$, and detector calibration constant $\beta$
with 3 seconds of data.  The initial spectral calibration step also yields the
mass $m_1$ with 3.0\% uncertainty. The subsequent equipartition method $m_2$
achieves an uncertainty of 4.1\% in 0.3 seconds.
\par The work presented here demonstrates the sensitivity of optical tweezers
in a scenario of strong environmental coupling, suggesting applications in
air-based sensing. For example, single-site ice nucleation could be monitored
in real time as a change in mass of the trapped particle. Alternatively, in a
system of constant mass, one could first measure the mass using the spectral
method and then use the equipartition method to measure changes in temperature
within the trapping medium, which could be driven out of equilibrium with a
temperature gradient to probe temperature-gradient-induced turbulence at small
scales of space and time.
\par The equipartition theorem may be challenged by non-equilibrium
dynamics. However, in the hydrodynamic regime where thermodynamic state
variables are relevant in the sense of quasi-equilibrium, we believe our
method will be quite applicable. The small sensor size means the dynamics are
fast to respond to changes in the environment (on the scale of $m/\gamma\sim
45~\si{\micro s}$ in this work). Even in the complete absence of thermal
equilibrium, where the notion of temperature is no longer defined, our position
and velocity data may be used to compute more general velocity structure
functions when the simple variance appearing in the equipartition theorem is
insufficient~\cite{kadanoff01,falkovich06}.  We consider such non-equilibrium
studies a fruitful direction for future optical tweezer experiments.
\section{Acknowledgments}
\par The authors are grateful to Y. Stratis, I. Bucay, Y. Lu, K. S. Melin,
S. Bustabad, L. Gradl  for helping with daily lab activities and making the lab
a pleasant environment. We also specially thank A. Helal for assisting with
SEM imaging of the silica microspheres.
\clearpage
\appendix
\section{\label{sec:parameters} Parameter conversions}
\par The physical parameters, denoted by the column vector
$\Params=\sups{(k,\rho,\beta)}{T}$, are functions of the independent variables
$\indeps = \sups{(a, b, c, R, \eta, T)}{T}$. First we define
\begin{gather}
    d_1 \equiv b + \sqrt{a c}\, , \\
    d_2 \equiv b + 2 \sqrt{a c}\, .
\end{gather}
%
% Note that the damping ratio, $\gamma / (2 \sqrt{k} m) = \sqrt{d_2 / (d_2-d_1)}/2$, is easily calculated with these definitions.
Then, for $\Params(\indeps)$, we have
\begin{gather}
    k(\indeps) = 12 \pi^2 \eta R \sqrt{\frac{a}{d_2}} \, , \\
    \rho(\indeps) = \frac{9 \eta}{4 \pi R^2} \sqrt{\frac{c}{d_2}} \, , \\
    \beta^2(\indeps) =  \frac{6 \pi^3 \eta R}{\kB T d_2}\, .
\end{gather}
\par The mass measurements $\vect{m} = \sups{(m_1, m_2, m_3)}{T}$ are a
function of the augmented independent variables, $\indeps^{\prime}=(a, b,
c, \mean{\dot{V}^2}, \mean{V^2}, R, \eta)$ (noting that $\pd{\vect{m}}{T} =
0$). For explicit formulae, we have
\begin{gather}
    m_1(\indeps^{\prime}) = 3 \eta R \sqrt{\frac{c}{d_2}} \, ,\\
    m_2(\indeps^{\prime}) = \frac{6 \pi^3 \eta R}{d_2} \frac{1}{\mean{\dot{V}^2}} \, ,\\
    m_3(\indeps^{\prime}) = 12 \pi^2 \eta R \sqrt{\frac{a}{d_2}} \frac{\mean{V^2}}{\langle \dot{V}^2 \rangle} \, .
\end{gather}
We next define
\begin{gather}
    u_1 \equiv \frac{3}{16 \pi^3 R^3}\, , \hspace{5mm}
    u_2 \equiv \frac{1}{2\sqrt{6 \pi \eta R \kB T}} \, , \\
    v_1 \equiv \frac{1}{4 \pi^2} \, ,\hspace{5mm}
    v_2 \equiv \frac{\pi}{\mean{\dot{V^2}}} \, , \hspace{5mm}
    v_3 \equiv \frac{\mean{V^2}}{\mean{\dot{V}^2}} \, ,
\end{gather}
to write the Jacobians
\begin{widetext}
\begin{gather}
    \label{eq:physical_jac}
     \PD{\Params}{\indeps} = \frac{6 \pi^2 \eta R}{\sqrt{a d_2^3}}
        \begin{pmatrix}
            d_1 & -a & -\sqrt{\Frac{a^3}{c}} & \Frac{2 a d_2}{R} & \Frac{2 a d_2}{\eta} & 0 \\
            -u_1 c & -u_1 \sqrt{a c} & u_1 d_1 \sqrt{\Frac{a}{c}} &  \Frac{-4 u_1 d_2 \sqrt{ac}}{R} &  \Frac{2 u_1 d_2 \sqrt{ac}}{\eta} & 0\\
            -u_2 \sqrt{c} & -u_2 \sqrt{a} & \Frac{-u_2 a}{\sqrt{c}} & \Frac{u_2 d_2 \sqrt{a}}{R} & \Frac{u_2 d_2 \sqrt{a}}{\eta} & -\Frac{u_2 d_2 \sqrt{a}}{T}
        \end{pmatrix}\, , \\
    \label{eq:mass_jac}
    \PD{\vect{m}}{\indeps^{\prime}} = \frac{6 \pi^2 \eta R}{\sqrt{a d_2^3}}
        \begin{pmatrix}
            -v_1 c & -v_1 \sqrt{a c} & v_1 d_1 \sqrt{\Frac{a}{c}} & 0 & 0 &  \Frac{2 v_1 d_2 \sqrt{a c}}{R} & \Frac{2 v_1 d_2 \sqrt{a c}}{\eta}\\
            -v_2 \sqrt{\Frac{c}{d_2}} & -v_2 \sqrt{\Frac{a}{d_2}} & \Frac{-v_2 a}{\sqrt{c d_2}} &
                \Frac{-v_2 \sqrt{a d_2}}{\mean{\dot{V}^2}} & 0 & \Frac{v_2 \sqrt{a d_2}}{R} & \Frac{v_2 \sqrt{a d_2}}{\eta}\\
            v_3 d_1 & -v_3 a & -v_3 \sqrt{\Frac{a^3}{c}} &
                \Frac{-2 v_3 a d_2}{\mean{\dot{V}^2}} & \Frac{2 v_3 a d_2}{\mean{V^2}} & \Frac{2v_3 a d_2}{R} & \Frac{2v_3 a d_2}{\eta}
        \end{pmatrix} \, ,
\end{gather}
\end{widetext}
which are needed for error propagation.
\section{\label{sec:system} Experimental Set Up}
\par Our experiment is sketched in Fig.~\ref{fig:fig1}  of the main
text and consists of a 1064 nm laser (Innolight Mephisto Laser System)
which is split into two counter-propagating, cross-linearly-polarized,
$\mathrm{TEM}_{00}$ beams. Each beam is passed through a 3 mm focal-length
aspheric lens (Thorlabs C330TMD-C) with foci offset by about $25~\si{\micro
m}$ along the optical axis. One beam is maintained at a higher power and
provides the confining potential at its focus while the counter-propagating
beam cancels the scattering force from the forward beam which otherwise ejects
the microsphere from the trap. The two counter-propagating beams thus form
a dual beam optical trap \cite{ashkin70,vanderhorst08}. An acoustic-optical
modulator (AOM) (IntraAction Corp, ATM-801A6 modulator) shifts the frequency
of the counter-propagating beam by 80 MHz which, in conjunction with the cross
polarization of the two beams, eliminates interference effects on the trapping
potential. Additionally, the AOM gives fine control over the power imbalance of
the two trapping beams. Silica microspheres (Bangs Laboratories, Inc, catalog
number SSD5001, nominal radius $1.5~\si{\micro m}$) are spread across a glass
coverslip which is fixed above the center of the trap and may be agitated with
a homemade piezoelectric transducer to release microspheres on demand.
\par The forward beam is isolated with a second polarizing beamsplitter
and its wave front is split with a D-shaped cut mirror (Thorlabs, model
BBD05-E03). Each half of the split beam is directed into the two ports of a
balanced photo detector (Thorlabs, model PDB120C, bandwidth DC-75 MHz), which
provides a voltage proportional to the optical power imbalance in the two
halves of the beam. The voltage signal is digitized (GaGe Razor 1622 Express
CompuScope) and saved to disk for analysis. When a microsphere is trapped,
the surrounding air causes it to undergo Brownian motion.  As the microsphere
deviates from the center of the trap, it scatters and deflects the trapping
light and the resulting signal from the balanced detector is proportional
to the displacement of the microsphere along the transverse direction
perpendicular to the cut of the mirror.  The forward beam's propagation
direction is taken as the $z$-axis and its polarization as the $y$-axis. The
cut mirror's edge is aligned to the $y$ axis and measures the displacement of
the particle along the $x$-axis. The entire setup is enclosed in a homemade
multi-chamber acrylic box \cite{bustamante09, nicholas14} to mitigate effects
of air currents. A thermocouple placed near the trap monitors the temperature
of the air.  Laser power is controlled with half wave plate and polarizing
beamsplitter pairs, one near the laser head to control the total trapping power
$P$, and one before the cut mirror maintains the total detected power at $\sim
35~\si{\micro W}$.
\section{\label{sec:methods} Methods}
\par Here we detail the methods of our data analysis.
\subsection{\label{sec:raw} Raw data}
\par The balanced photodetector has two ports, $a$ and $b$, for measuring
optical power.  The device provides three continuous voltage signals,
$\subs{V}{a}(t)$, $\subs{V}{b}(t)$, and $V_{-}(t) = \subs{V}{b}(t) -
\subs{V}{a}(t)$ with a bandwidth of 0 to 75 MHz. Once a microsphere is
trapped and just before an experimental trial begins, we make a measurement of
$V_{+}(0) = \subs{V}{a}(0) + \subs{V}{b}(0)$. We then make a digitized record
of $V_{-}(t)$ consisting of $N=2^{24}\approx 1.7 \times 10^7$ samples at a
rate of $r = 5 \times 10^7$ Hz. Hence, one trial is of length $\mathcal{T} =
N/r = 0.336$ s. We collect 10 trials for a total of 3.36 s of data at each of
14 different trapping laser powers. Note that while $V(t)$ is adimensional, we
refer to it as the \emph{voltage signal}. Similarly $\dot{V}(t)$ is referred to
as \emph{voltage-derivative} signal.
\subsection{\label{sec:averaging} Bin-averaging}
\par Post processing begins by, for each trial, forming a data set
$V = \{V_j\}$, where $V_j = V_{-}(t_j) / V_{+}(0)$ is the signal at
time $t_j = j/r$, for $j=0, 1, \dots , N-1$. Next, we average $V_j$ over
non-overlapping bins of size $M=2^8$ to form $\mean{V}(M_j)$, where $M_{j}$
is the $\sups{j}{th}$ time bin and $j=0,1,\dots, \left(N/M\right) - 1$. The
bin-averaging procedure reduces uncorrelated voltage noise by a factor of
$M^{-1/2}$, and decreases the sampling rate and number of points both by
a factor of $M^{-1}$. Explicitly, the $\sups{j}{th}$ $M$-bin average of a
digitized quantity $q$ is
\begin{equation}
    \mean{q}(M_j) = \frac{1}{M} \sum_{i=0}^{M-1} q_{j M + i} \,,
    \label{eq:binaverage}
\end{equation}
where $q_{j M + i}$ is the $\sups{(jM+i)}{th}$ element of $q$. For our
choice of $M=2^8$, the number of data points becomes $N/M = 2^{16} \approx
6.6\times 10^4$. The new effective sampling rate, $r/M = 195~\si{kHz}$, is
still roughly 10 times faster than the momentum decay rate $\gamma/m \approx
22~\si{kHz}$ which sets the time scale of correlated ballistic motion. Since
a 256-bin average is always the first step in our data processing, we
reuse the notation $V=\{V_j\}$ to denote a trial after processing with
Eq.~\eqref{eq:binaverage}. Additionally, we take $N$ and $r$ to henceforth
denote the values after bin averaging. The instantaneous velocity may now be
meaningfully computed.
\subsection{\label{sec:velocity} Instantaneous velocity}
\par The instantaneous velocity of the microsphere is proportional to the
voltage-derivative signal. We approximate the derivative numerically with the
$8^{\mathrm{th}}$-order central finite difference approximation
\begin{equation}
\begin{split}
    \dot{V}_j = r \bigg (
    &\frac{V_{j-4}}{280} - \frac{4 V_{j-3}}{105} +
    \frac{V_{j-2}}{5} -\frac{4 V_{j-1}}{5} + \\
    &\frac{4 V_{j+1}}{5} - \frac{V_{j+2}}{5} +
    \frac{4 V_{j+3}}{105}  - \frac{V_{j+4}}{280} \bigg )
    + \mathcal{O}(r^{-8}) \, .
    \label{eq:firstdiff}
\end{split}
\end{equation}
With this approximation, we can compute a total of $N - 8$ velocities
corresponding to the times $t_j^{(\dot{V})} = t_j$ for $j=4,5,\dots,N-5$.
\subsection{Signal variance}
\par The voltage signal variance $\mean{V^2}$ and voltage-derivative signal
variance $\mean{\dot{V}^2}$ are needed for mass measurement methods $m_2$ and
$m_3$. For the quantities $q = V, \dot{V}$, we fit the density histogram of
$\{q_j\}$ to a normalized one-dimensional Gaussian distribution
\begin{equation}
    \subs{\mathcal{P}}{G_1}(q;\mean{q^2}) = \frac{1}{\sqrt{2 \pi \mean{q^2}}}
        \exp \left (\frac{-q^2}{2 \mean{q^2}} \right ) \, ,
\end{equation}
for the free parameter $\mean{q^2}$. The fits are performed as un-weighted
chi-squared minimizations. We take $L=10$ trials of each data set. The
statistical variance of $\mean{q^2}_i$ across trials $i=0,1, \dots L-1$
relative to the mean value
$\mu_{\mean{q^2}} = \frac{1}{L}\sum_{i} \mean{q^2}_i$ is
\begin{equation}
    \sigma^2_{\mean{q^2}} =
        \frac{1}{L} \sum_{i=0}^{L-1}
            \left ( \mean{q^2}_i - \mu_{\mean{q^2}} \right )^2 \, .
    \label{eq:var-stat-var}
\end{equation}
The square-root of equation~\eqref{eq:var-stat-var} gives the uncertainty in
$\mean{q^2}$ for the purposes of error propagation.
\par Both $\mean{\dot{V}^2}$ and $\mean{V^2}$ are sensitive to the amount of
initial bin averaging. The choice of $M=256$ was arrived at by converging
the two variances as a function of $M$. Too little averaging and the
voltage-derivative variance is over-estimated. Too much averaging and both
variances are under-estimated. $M=256$ is between these two extremes where the
change in variance is minimal for a small change in $M$.
\par The Allan deviation, is the square-root of the Allan variance (two-sample
variance)~\cite{allan66},
\begin{equation}
    \mathcal{A}^2_{\mean{q^2}}(\tau) =
        \frac{1}{A-1} \sum_{j=0}^{A-2} \frac{1}{2}
            \left [ \mean{q^2}(\tau_{j+1}) -  \mean{q^2}(\tau_j) \right ]^2\, ,
\end{equation}
where $A = \mathcal{T} / \tau$ is the number of independent length-$\tau$
blocks in a trial of total length $\mathcal{T}$ and $\mean{q^2}(\tau_j)$ is the
$\sups{j}{th}$ bin-average of $q^2$ with $M = r \tau$ points.
\subsection{\label{sec:bartlett}
    Voltage power spectral density: Bartlett's method}
\par Next, we describe our method of extracting the experimental voltage
PSD. The one-sided PSD calculated from a single trial $\{V_j\}$ is
\begin{gather}
    \label{eq:psdcalc}
    \hat{S}_{V,k} = \frac{2}{W_N}
                     \left \lvert \sum_{j=0}^{N-1} w_j^{(N)} V_j
                     \exp \left [ -2\pi i \frac{j k}{N} \right ] \right \rvert^2 \, ,\\
    \label{eq:psdcalc_freq}
   f_k = \frac{k r}{N} \, ,
\end{gather}
for $k=0,1,\dots, N/2$. We have introduced the notation $\hat{S}_{V,k} =
\hat{S}_V(f_k)$. Additionally, $w_j^{(N)} = w(t_j; N)$ is a windowing function
for a trial of size $N$ and
\begin{equation}
    W_N = \frac{1}{r}\sum_{j=0}^{N-1} \left \lvert w_j^{(N)} \right \rvert^2 \, ,
\end{equation}
is a power correction for the windowing procedure. The factor $2/W_N$ ensures
Parseval's theorem $\sum_k \hat{S}_{V,k} = \mean{V^2}$. Note that if $w(t_j; N)
= 1$, then $W_N = \mathcal{T}$. We use the a Hamming window
\begin{equation}
    w(t_j; N) = 0.54 - 0.46 \cos \left ( \frac{2 \pi t_j}{t_{N-1}} \right ) \, ,
\end{equation}
for which $W_N = 0.4194 \mathcal{T}$ in the continuum limit $N \rightarrow
\infty$ and $r \rightarrow 0$. The choice of window was found to be
inconsequential for parameter extraction except for small deviations at the
lowest few trapping-laser powers, i.e., the over-damped oscillator regime.
This observation appears to be in analogy with~\cite{dawson19} where Bartlett's
method is used in conjunction with windowing because of the accuracy it affords
parameter extraction at high vacuum pressure, i.e., high damping.
\par $V_j$ is proportional to the microsphere's position at time $t_j$, and the
microsphere is subject to a stochastic force $F(t_j) \propto \xi(t_j)$, where
$\xi(t_j)$ is a real-valued, Gaussian distributed, white noise random process
with zero mean and delta-correlation. A single trial's PSD $\hat{S}_{V,k}$,
then, is subject to fluctuations $\propto \lvert \tilde{\xi}(f_k) \rvert
^2$ where $\tilde{\xi}(f_k)$ is the discrete Fourier transform of the
particular instance $\xi(t)$ present in that trial. Since $\xi(t)$ is
Gaussian-distributed, $\tilde{\xi}(f_k)$ is a complex number with real and
imaginary parts independent and identically Gaussian distributed. $\lvert
\tilde{\xi}(f_k) \rvert ^2$ is the sum of two squared Gaussian-distributed
variables which is exponentially distributed~\cite{berg-sorensen04}. Thus,
the experimental $\hat{S}_{V,k}$ is an exponentially distributed random process
with mean and standard deviation equal to the theoretical PSD, $S_{V,k}$:
\begin{equation}
    \mathcal{P}(\hat{S}_{V,k}) =
        \frac{1}{S_{V,k}} e^{- \hat{S}_{V,k}/ S_{V,k}}
\end{equation}
It is desirable to suppress this intrinsic noise so that fitting experimental
data yields precise parameters.
\par The idea of Bartlett's method~\cite{oppenheim01} is to average several
independent noisy PSDs. First, the time-domain signal is segmented into $n$
independent blocks. Then, for each block we calculate a PSD, and average the
results. To proceed, we form $V^{(\ell)}_j$ where $\ell = 0, 1, \dots, n - 1$
enumerates blocks of size $N/b$ given by partitioning each of the $L=10$ trials
of size $N$ into $b=4$ blocks. Hence, there are a total of $n = L b = 40$
blocks of size $N/b = 2^{14}$ (or length 83.9~\si{ms}). Bartlett's PSD estimate
may be written
\begin{gather}
    \label{eq:psdbartlett}
    \hat{\bar{S}}_{V,k} = \frac{1}{n}
                     \sum_{\ell=0}^{n - 1} \hat{S}_{V^{(\ell)},k} \\
    \label{eq:psdbartlett_freq}
   \bar{f}_k = \frac{k r b}{N} \, ,
\end{gather}
where it is understood that here the $N$ appearing in Eq.~\eqref{eq:psdcalc}
is the size of $V^{(\ell)}$, namely $N/b$.
\par Due to the PSD averaging, $\hat{\bar{S}}_{V,k}$ is
\emph{Gamma} distributed (the convolution of $n$ exponential
distributions)~\cite{norrelykke10}:
\begin{equation}
    \mathcal{P}(\hat{\bar{S}}_{V,k}) = \frac{1}{S_{V,k}} \frac{n^n}{\Gamma(n)}
        \left ( \frac{\hat{\bar{S}}_{V,k}}{S_{V,k}} \right )^{n-1}
         \exp \left (-n \frac{ \hat{\bar{S}}_{V,k}}{S_{V,k}} \right )\, ,
     \label{eq:gamma2}
\end{equation}
where $\Gamma(n) =(n-1)!$ is the gamma function.
\par The probability of measuring the model $S_{V,k}=S_V(f_k,\params)$ given
the data $\hat{\bar{S}}_V$ is the product over $k$ of Eq.~\eqref{eq:gamma2}, i.e.,
\begin{equation}
    \mathcal{P}(\hat{\bar{S}}_V | \params) =
        \prod_{k=0}^{N/2-1} \mathcal{P}(\hat{\bar{S}}_{V,k}).
    \label{eq:likelihood2}
\end{equation}
Henceforth and in the main we simplify the notation and take
$\hat{S}_V(f_k)$ to denote the Bartlett PSD discussed here. Optimal
parameters $\hat{\params}$ are found by maximizing Eq.~\eqref{eq:likelihood2},
or minimizing its negative-logarithm $\mathcal{L}(\hat{S}_V, \theta)$
(Eq.~\eqref{eq:objective} of the main text).
\section{\label{sec:uncertainty} Uncertainty analysis}
\par In this section, we elaborate on the quantification of uncertainty in
voltage measurements, fitting parameters, and fixed parameters.
\subsection{\label{sec:voltage} Voltage measurement uncertainty}
\par The signal digitizer has 16 bits and a full-scale input range
of 2 $V$ ($\pm 1~\si{V}$).  However, the manufacturer reports an
effective-number-of-bits of 11.7 (meaning 4.3 bits are polluted by electronic
noise). The resolution of the digitizer is the voltage range allotted to the
least significant bit (LSB): $\mathrm{LSB} = 2 V /2^{11.7} = 601~\si{\micro
V}$. The systematic uncertainty in the measured voltage $V_{-}$ due to
digitization is estimated to be $\sigma_{V_{-}} = \frac{1}{2}\mathrm{LSB}$.
In the experiments reported on here, we need to resolve voltage variances. The
smallest observed standard deviation is $\mean{V^2_{-}}^{1/2} = 0.02~\si{V}$,
so a worst-case estimate of the systematic uncertainty due to digitization is
about 1.5\%.  Upon bin-averaging with $M=256$ points, the standard deviation of
a purely electronic-noise signal is found to be reduced by the expected factor
$M^{-1/2} = 1/16$ (which we have verified with experimental data). Thus, the
systematic uncertainty in voltage measurements due to this electronic noise is
$\sigma_{V_{-}} / \mean{V^2_{-}}^{1/2} = 0.09\%$, which is ignorable compared
to other uncertainties.
\begin{figure}
\includegraphics[width=3.2in]{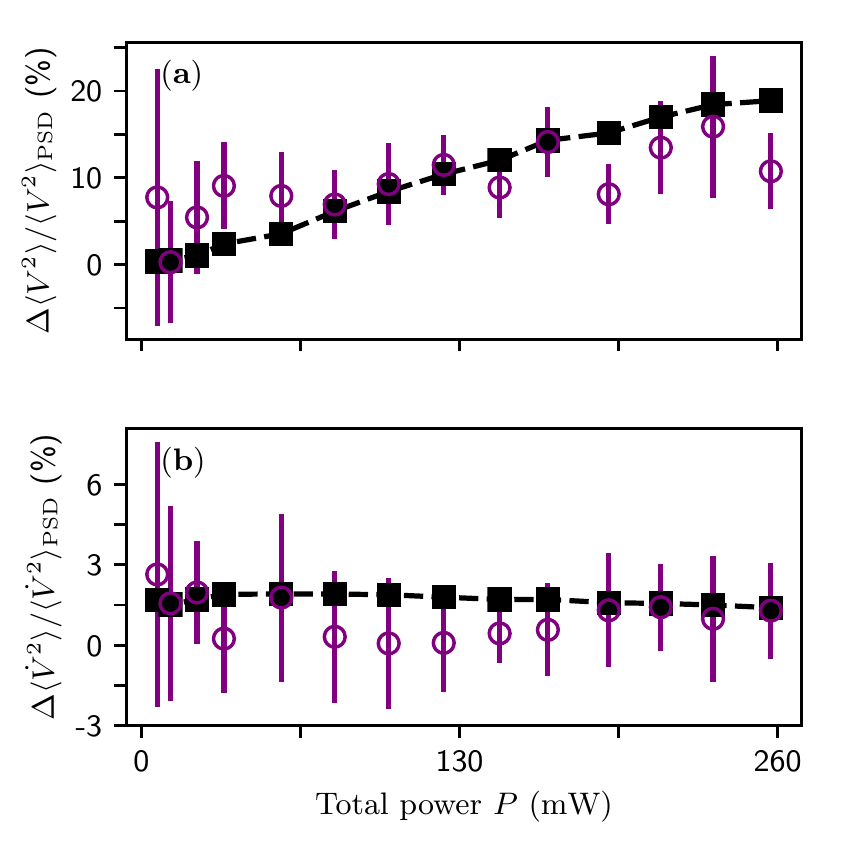}
    \caption{\label{fig:fig7}
    (a) Excess variance is observed in the time domain voltage signal
        as compared to the variance expected by fitting the PSD (purple
        circles). The black squares reflect the integral of the noise spectrum
        over the 120 Hz peak under the same normalization.
    (b) Excess variance observed in the voltage-derivative signal (purple
        circles) and the expected excess variance based in the integral of the
        noise spectrum from 80 kHz and above. In both panels the dashed line
        is a guide to the eye, and error bars reflect the statistical variation
        of variance measurements across 10 trials for each of the 14 different
        trapping laser powers.
}
\end{figure}
\par The above discussion considers only electronic noise of the detector
and digitizer. However, uncertainty in a measured signal variance is much
more significantly affected by laser noise, which includes pointing and power
fluctuations. Pointing instability is inherent to the laser and also arises
due to mechanical and electrical coupling through the the mirrors and control
electronics. The significant resonance observed near 120 Hz in the laser-on
noise spectrum (Fig.~\ref{fig:fig2}(a) of the main text) accounts well
for the excess variance observed in the time domain voltage signal as compared
to the variance expected based on PSD fitting, which easily excludes the
noise peak in the fitting procedure. In Fig.~\ref{fig:fig7}(a) we show the
observed excess variance $\Delta\langle V^{2}\rangle = \lvert \mean{V^2} -
\subs{\mean{V^2}}{PSD}\rvert$ as well as the value predicted by integrating
the noise spectrum in a band from 70 Hz to 170 Hz. On average, the systematic
uncertainty in $\mean{V^2}$ due to this excess variance is at about 10\%, which
seemingly explains the systematic over estimate of $m_3$ (which is proportional
to $\mean{V^2}$) compared to $m_1$ and $m_2$.
\par The voltage-derivative signal is insensitive to the presence of the 120
Hz resonance. Instead, the observed excess variance is well accounted for high
frequency noise (above 80 kHz). Figure \ref{fig:fig7}(b) shows the observed and
expected excess variance in the voltage-derivative for each set of trials. On
average, the systematic uncertainty in $\mean{\dot{V}^2}$ is about 2\%.
\subsection{Uncertainty in PSD fitting parameters}
\par
\begin{figure}[t]
\includegraphics[width=3.375in]{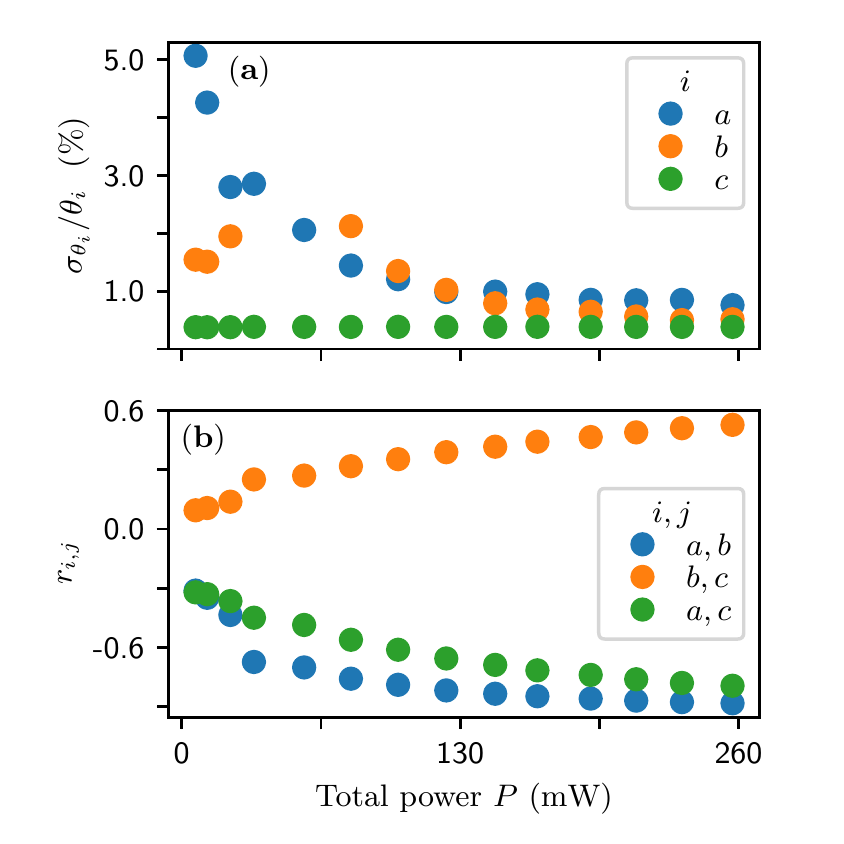}
\caption{ \label{fig:fig8}
    (a) Relative uncertainty in fitting parameters and (b) correlation
    coefficients at different total trapping laser powers. The parameter $b$
    goes through a zero crossing near 65 mW causing the relative uncertainty
    to diverge. We omit two data points in that region for visualization
    purposes.}
\end{figure}

\begin{figure}[t]
\includegraphics[width=3.2in]{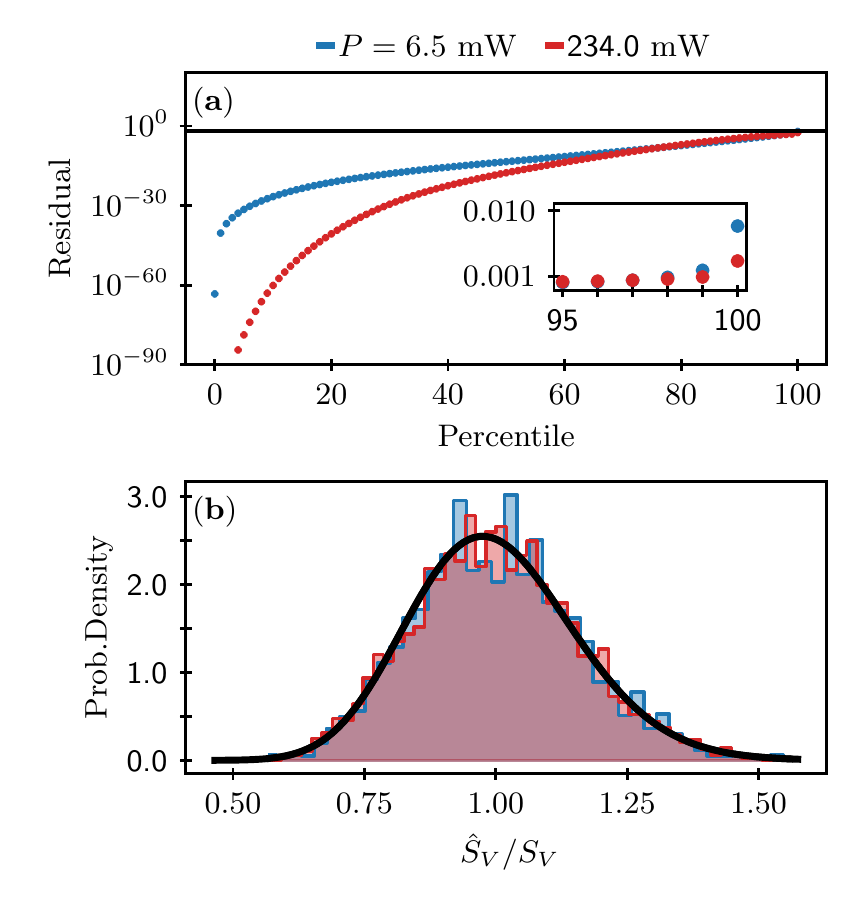}
\caption{ \label{fig:fig6}
    (a) Absolute residual between the fitting-parameter-scan data and
        three-dimensional Gaussian fit are bounded by 1\% (black line). The
        inset demonstrates the $95^{\mathrm{th}}$ percentile is bound by 0.1\%.
    (b) Distribution of the ratio between PSD data and PSD fit (filled
        step-plots) and the expected gamma distribution for $n=40$ PSD averages
        (no free parameters, black line).}
\end{figure}

 The PSD fitting parameters $\params =\sups{(a,b,c)}{T}$ and their
associated variance-covariance matrix $\vect{\Sigma}_{\params}$ are found by
calculating Eq.~\eqref{eq:likelihood2} in the vicinity of the optimal parameters
(main text Fig.~\ref{fig:fig2}(b)-(c)) and fitting the result to a
three-dimensional Gaussian. For completeness, the relative uncertainties
and correlation coefficients of the fitting parameters are plotted in
Fig.~\ref{fig:fig8}. We find excellent agreement between the three-variate
data and fit in Fig.~\ref{fig:fig6}(a) which shows the absolute residual of the
fit from the data as a function of percentile for high and low trapping laser
powers.
\par The optimal parameters generate a model $S_V$ around which the
PSD data $\hat{S}_V$ is scattered according to the gamma distribution
\eqref{eq:gamma2}. Hence, distribution of $\hat{S}_V/S_V$ is also
gamma-distributed, but with unit mean. In Fig.~\ref{fig:fig6}(b) we show
the observed distribution of $\hat{S}_V/S_V$ for high and low trapping laser
powers as well as the theoretically expected gamma distribution for $n=40$
PSD averages.
\begin{figure*}[!ht]
\includegraphics[width=0.9\textwidth]{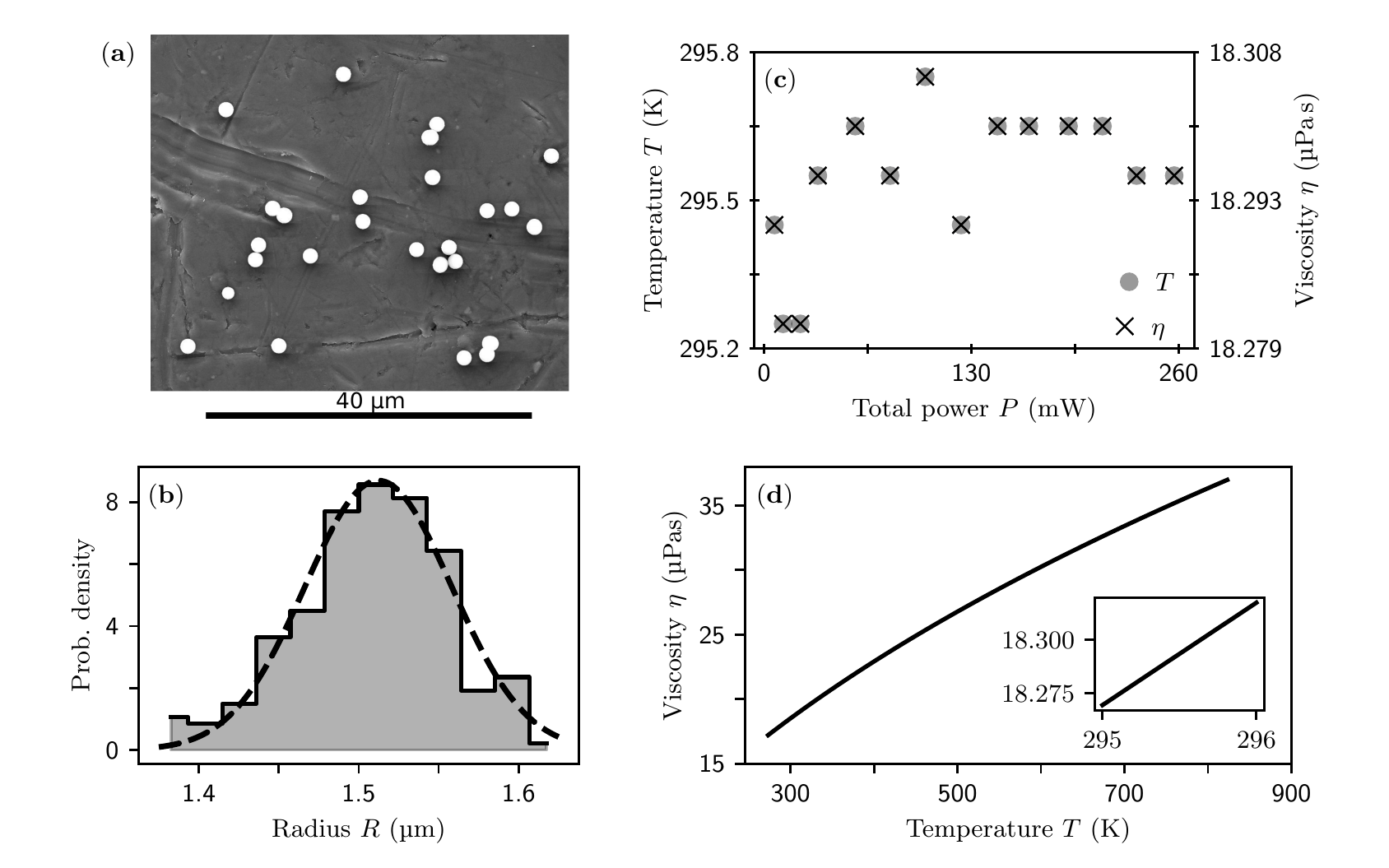}
    \caption{ \label{fig:fig5}
             (a) Example SEM image analyzed for the microsphere radius.
             (b) A density histogram (filled gray step-plot) of the 219
                 observed radii. The dashed line is a Gaussian fit with
                 mean and variance as free parameters.
             (c) Observed temperatures (circles) and the corresponding
                 viscosity (crosses) evaluated with the Sutherland model
                 for each set of the 14 sets of trials.
             (d) Temperature dependence of viscosity according to the
                 Sutherland model (Eq.~\eqref{eq:eta}). The inset shows
                 the linearity of viscosity over the experimentally relevant
                 temperature range.
}
\end{figure*}
\subsection{\label{sec:fixed_params} Uncertainty in fixed parameters}
\par To fix the microsphere radius, we first took 10 scanning electron
microscope (SEM) images of a sample of $3~\si{\micro m}$ diameter
microspheres. An example image is shown in Fig.~\ref{fig:fig5}(a). We
measured the diameter of the imaged spheres using the Particle Sizer
plugin~\cite{particlesizer} for the Fiji distribution of ImageJ, an open source
image analysis software. We operated the plugin with default settings and in
``single particle mode" to exclude touching and overlapping microspheres.  Once
an image was analyzed, we manually excluded false positives by inspection. In
total, we measured 219 spheres and fit the histogram of measured radii
to one-dimensional Gaussian probability density (Fig.~\ref{fig:fig5}(b)).
The fitting procedure yields a mean of $R=1.512~\si{\micro m}$. The uncertainty
of fixing $R$ to this value is taken to be the standard deviation of the fitted
distribution, $\sigma_R=0.044~\si{\micro m}$. Our measurement provides a marked
improvement over the manufacturer-stated radius of 1.5~\si{\micro m} with an
uncertainty of 10\%.
\par The temperature $T$ is measures at the beginning of each set of trials
using type-K thermocouple. Over the entire experiment reported on here,
the temperature varied by only 0.05\%. As a result, we take $T$ of a set of
trials to be given by the value measured at the start of collection and the
uncertainty is taken as $\sigma_T=0$ in comparison to other uncertainties.
\par The Dynamic viscosity $\eta$ as a function of temperature $T$ is
evaluated with the Sutherland model~\cite{chapman90,mulholland06} (plotted in
Fig.~\ref{fig:fig5}(d))
\begin{equation}
    \eta(T) = \eta^{\prime} \left ( \frac{T}{T^{\prime}} \right )^{3/2}
              \frac{T^{\prime} + S}{T + S} \, ,
    \label{eq:eta}
\end{equation}
in which $\eta=\eta^{\prime} = 18.3245~\si{\micro Pa}$ when
$T=T^{\prime}=296.15~\si{K}$, and $S=110.4~\si{K}$ is called Sutherland's
constant. $S$ is roughly a measure of the mutual potential energy in a system
of two air molecules in contact. Over such a small temperature range, viscosity
is linear (inset of Fig.~\ref{fig:fig5}(d)). We set $\eta$ for each set of
trials to be the value given by Eq.~\eqref{eq:eta} evaluated at that trail's
measured temperature. The variation across the entire experiment was found to
be 0.04\% so we also set $\sigma_{\eta}=0$.
\subsection{\label{sec:propagation} Propagation of errors}
\par The variance-covariance matrix of the physical parameters is given
by $\vect{\Sigma}_{\Params} = \vect{J}_{\Params} \vect{\Sigma}_{\indeps}
\sups{\vect{J}}{T}_{\Params}$.  where $\vect{\Sigma}_{\indeps} =
\mathrm{diag}(\vect{\Sigma}_{\params}, \sigma^2_R, \sigma^2_{\eta},
\sigma^2_T)$ and the Jacobian $\vect{J}_{\Params}$ is evaluated at
$\hat{\params}$ using Eq.~\eqref{eq:physical_jac} of the main
text. Simmilarly, the variance-covariance matrix of the mass measurements
if given by $\vect{\Sigma}_{\vect{m}} = \vect{J}_{\vect{m}}
\vect{\Sigma}_{\indeps^{\prime}} \sups{\vect{J}}{T}_{\vect{m}}$ where
$\vect{\Sigma}_{\indeps^{\prime}} =
    \mathrm{diag} \left ( \vect{\Sigma}_{\params}, \sigma^2_{\mean{\dot{V}^2}},
    \sigma^2_{\mean{V^2}}, \sigma^2_R, \sigma^2_{\eta} \right )$.
The uncertainty in any parameter is then given by the square root of the
corresponding diagonal element of the appropriate variance-covariance matrix.
\par Similarly, correlations between parameters are quantified in the off
diagonal elements of the appropriate variance-covariance matrix. Correlations
between physical parameters are expected because both the calibration constant
and trap strength depend on the index of refraction of the microsphere,
which is correlated with its density. It is correlations between the fitting
parameters that we are careful to account for because they contribute to the
uncertainties of derived parameters.
\clearpage
\bibliography{weighing} % Produces the bibliography via BibTeX.
\end{document}